\documentclass[prb,preprint,letterpaper,noeprint,longbibliography,nodoi,footinbib]{revtex4-1} 

\usepackage[colorlinks, allcolors=blue]{hyperref}
\usepackage{amsmath}  
\usepackage{amsfonts} 
\usepackage{graphicx} 
\usepackage[utf8]{inputenc} 
\usepackage{indentfirst}
\usepackage{tikz}
\usepackage{amssymb}
\usepackage{hyperref}       
\usepackage{url}            
\usepackage{booktabs}       
\usepackage{amsfonts}       
\usepackage{nicefrac}       
\usepackage{microtype}      
\usepackage{lipsum}
\usepackage{bbm}

\begin{document}


\title{On the Laws of Thermodynamics in Classical Non-Inertial Frames}

\author{Wendel M. Mendes}
\email{wendel.mendes@ifce.edu.br} 
\altaffiliation[permanent address: ]{Direção de Ensino, Instituto Federal do Ceará, Campus Crateús,
  Av. Geraldo Marques Barbosa, 567 - Venâncios CEP: 63.708-260. Crateús - CE Brasil} 
\author{Bruno P. e Silva}
\email{bruno.poti@ifce.edu.br}
%


\date{\today}

\begin{abstract}

This work consists in the theorical development on the analysis of the Thermodynamic Laws and thermodynamic systems in relative motion, according to the laws of Classical Mechanics. The difference of this work for many of the literature is the methodology used. We follow a strictly mathematical line, instead of a phenomenological line. We proved that the First and Second Laws of Thermodynamics in non-inertial frames remain invariant under any spatial translation between reference frames in motion. From the results obtained, we show the transformation of the main thermodynamic quantities as temperature, pressure, chemical potential, internal energy, heat, state equation for an ideal gas and efficiency of Carnot's machine are also invariant under any transformation between non-inertial reference frames.
\end{abstract}

\maketitle

\section{Introduction}

The problem of motion remotes to the beginning of humanity civilization and was certainly one of the first problems formulated by man, resulting in the birth of \textit{Classical Mechanics}, that is a branch of physics in which studies the motion of material bodies and your respective causes \cite{Noel}. For a long time, it was thought that motion was an absolute concept, regardless of the point at which it was observed. However, in the course of human history, it was found that the motion was a \textit{relative} concept, that is, it depends on the observer \cite{Werner}.

A systematic and rigorous study of Classical Mechanics requires precise definitions of \textit{space} and \textit{time}. The space is one of the most primitive concepts used by the man and is found operationally defined by distance measures, angles, area and volume that where introduced by \textit{euclidean geometry}, written by Euclides through his work titled \textit{Elements}. The time is also an old and primordial concept in the study not only of Classical Mechanics, but of Physics in general. Physically it's more important to define time operationally, i.e, in terms of measures. This was the definition of Galileo Galilei when he discovered the isochronism of the oscillation of pendulum when he compared the oscillations of the Pisa Cathedral with his pulse rhythm. This led to the conclusion that measures of time is related to periodic phenomena \cite{Noel}.

An attentive reader can see that the formulation of the Classical Thermodynamic theory, systematically formulated in the 19th and 20th centuries don't mention anything about the transformation of thermodynamics quantities when measured in different reference frames that are in relative motion \cite{Callen}. Most, if not all, of the textbooks on this subject, consider only systems that are at rest in relation to the observer, making no relation to galilean relativity. Some of the first to address the relation between thermodynamics and galilean relativity was the paper of Tykodi \cite{Tykodi} that is extremely phenomenological and deals only with inertial references and as well as the article of Van Kampen \cite{Van}, that focuses on Einstein special relativity and thermodynamics and Relativity, but contains a very small section only about the first law of Thermodynamics and newtonian classical relativity. The Authors of this work noted a scarcity of didactic papers related to Physics on this subject. Many are linked to Engineering, in the context of \textit{Extended Thermodynamics}, which is a theory that complements the usual laws of conservation of mass, momentum and energy, employed in the Classical Thermodynamics \cite{Muller,Ruggeri}. 

Before we approach thermodynamics in the context of transformation between different reference frames, a brief review of the main concepts about the relativity of motion is presented in the section \ref{Newton} and the main aspects of the equilibrium Thermodynamics in the section \ref{Termo}. The actual subject of this paper begins in the section \ref{Transf}, in which we approach the laws of Thermodynamics in accelerated reference frames, from microscopic and macroscopic point of view, and the transformation of thermal quantities in both reference frames. After exposing the principles, in the section \ref{Aplic}, two applications: (i) barometric differential equation for an ideal gas and (ii) Carnot thermal machine.

\section{Review of Coordinate System Motion}
\label{Newton}

The concepts of Classical Mechanics launched by Newton are not clear in some point. An example of this is linked to the fact that no specification is made about the \textit{coordinate system}, in which the accelerations, mentioned in the two Newton's laws, must be measured \cite{Symon}. Newton recognized this problem and this was only better understood later through the experimental facts that showed the validity of the laws of mechanics in any moving reference frame system in relation to other coordinate systems with constant velocity. This is called \textit{Newtonian Principle of Relativity}, which will be covered in this section.

A \textit{physical event} is a fundamental concept used to study the relativity of motion according to the laws of the Classical Mechanics. So, a physical event can be defined as a phenomenon that occurs regardless of the choice of reference frame, in a given position $\mathbf{r}$ of three-dimensional space. Examples of physical events are the collisions between two particles or the turn on - turn off of light source. All this occurs independently in relation to the choice of frame of reference.

It's known that the notion of rest or motion in Classical Mechanics depends on the frame adopted. Also, it can be noticed that the idea of rest or movement depends on a real parameter called \textit{time}. Consider two arbitrary observers $O$ and $O'$. It is an experimental fact at the limits of Classical Mechanics that establishes that all observers measures the same time interval of the occurrence of a given event $P$, that is, if one observer is in an instant of time $t$ and another is in an instant of time $t'$, both measure the same time of occurrence of $P$, so  \cite{Landau}
\begin{eqnarray}
       t'=t.
\end{eqnarray}
On the other hand, it is known fact in Classical Mechanics that the behavior of the measuring instruments is not affected by the state of motion. Here we can note the main distinction between Classical Mechanics and Quantum Mechanics \cite{Costas}. This distinction stems from the fact that it’s possible to isolate a physical object to make a measure, what is not possible at the quantum limits due to the Heisenberg's uncertainty principle \cite{Gasi}.

Another relevant aspect in the discussion of relativity in Classical Mechanics concerns the velocity of interactions. Any interaction between two macroscopic bodies occurs  instantly, regardless of how apart they are. We are soon led to conclude that the velocity of interaction is infinite. This stems from the fact that in order to observe or measure any physical quantity it is necessary to use or handle light. It has a propagation speed in the vacuum higher than any macroscopic object, so it is reasonable to assume that the speed of light is limitless in Classical Mechanics. 

The space where the physical events must occur must be homogeneous and isotropic \cite{Tung}. The spatial and temporal homogeneity refers to the fact that the mechanical phenomena are independent of the place and of the instant of time, respectively, from where they are observed or measured, while that the isotropy refers to the independence of the orientation of isolated physical systems. From these symmetries, important conservation laws of Classical Mechanics can be extracted, such as conservation of linear momentum, energy and angular momentum
 \cite{Tung,Jerrold}.

Now let's consider the kinematic description of a given physical event situated in $\mathcal{P}$, from the point of view of two referential frames. Consider that one of referential frames has its origin $\textbf{O}$ and other in $\textbf{O'}$, called \textit{mobile referential}, which moves in relation to \textbf{O} with constant velocity \textbf{u}. 

\begin{figure}[h]
\center
\begin{tikzpicture}
\draw[thick][->] (0,0,0)--(2,0,0);
\draw[thick][->] (0,0,0)--(0,2,0);
\draw[thick][->] (0,0,0)--(0,0,2);
\draw node (nO) at (0.1,-0.3,0) {$O$};
\draw node (n1) at (2,-0.5,0) {$x$};
\draw node (n2) at (-0.5,2,0) {$y$};
\draw node (n3) at (0.5,0,2.5) {$z$};
\draw[thick][->] (4,2.5,3)--(6,2.5,3);
\draw[thick][->] (4,2.5,3)--(4,4.5,3);
\draw[thick][->] (4,2.5,3)--(4,2.5,5);
\draw node (nO') at (4.1,2.2,3) {$O'$};
\draw node (n1') at (5,1,0) {$x'$};
\draw node (n2') at (2.5,3.5,0) {$y'$};
\draw node (n3') at (3.5,1.5,2.5) {$z'$};
\draw[ultra thick][->] (0,0,0)--(4,2.5,3);
\draw node (h) at (1.5,0.4,0) {$\mathbf{h}$};
\draw[ultra thick][->] (0,0,0)--(3.6,3.1,0);
\draw node (r) at (1.8,2,0) {$\mathbf{r}$};
\draw[ultra thick][->] (4,2.5,3)--(3.7,3,0);
\draw node (r') at ((4.7,3.2,3) {$\mathbf{r'}$};
\draw node (ponto1) at (3.7,3.1,0) {\textbf{.}};
\draw node (P) at (4,3.1,0) {$P$};
\end{tikzpicture}
\caption{Scheme of Galileo's Transformations.}
\label{referenciais}
\end{figure}
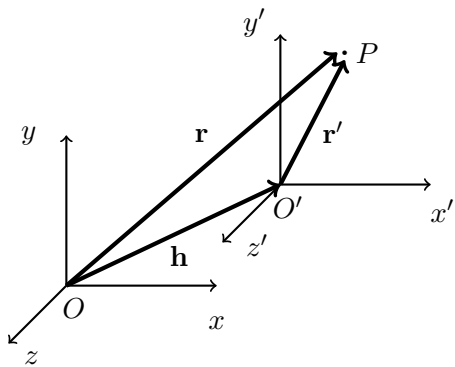

For the situation presented in the Fig. \ref{referenciais}, the vectorial equation between the vectors $\mathbf{r}$ and $\mathbf{r}$' is
\begin{eqnarray}
    \textbf{r} = \textbf{r}' + \textbf{h},
\label{r(t)}
\end{eqnarray}
where $\mathbf{r}$ is the measure of position of $P$ with respect to the origin $O$ and $\mathbf{r}$' is the measure of position of $P$ of respect to the origin $O'$. If the origin $O'$ moves in relation to $O$, then
\begin{eqnarray}
    &\textbf{v}&=\frac{d\textbf{r}}{dt} = \frac{d\textbf{r}'}{dt'} + \frac{d\textbf{h}}{dt} = \textbf{v}' + \textbf{u},
\label{advel}
\end{eqnarray}
where $\displaystyle \textbf{v}' = \frac{d\textbf{r}'}{dt'}$ is the velocity of point $P$ relative to $O$' and $ \displaystyle \textbf{v} = \frac{d\textbf{r}}{dt}$ is the velocity of point $P$ relative to $O$. We can define the velocity of $O'$ relative to $\textbf{O}$ for $\displaystyle \textbf{u} = \frac{d\textbf{h}}{dt}$. Note that, since we assume that the time is absolute, we have $\displaystyle \frac{dt'}{dt}$ = 1. The equation (\ref{advel}) is called \textit{Galileo's velocity addition}. If the movement of $O'$ in relation to $O$ is such for every moment of time $t$, the Cartesian axes of the $O$ and $O'$ referential frames are always parallel to each other, we may say that the displacement \textbf{h}(t) of $\textbf{O}$' is called \textit{coordinate system translation} $O'$

The relation between the relative accelerations is then easily deductible, so that we have
\begin{eqnarray}
   \mathbf{a} = \frac{d\mathbf{v}}{dt} = \frac{d\mathbf{v}'}{dt} + \frac{d\mathbf{u}}{dt}= \mathbf{a}' + \mathbf{a}_u,
   \label{adacel}
\end{eqnarray}
where $\displaystyle \textbf{a}=\frac{d\textbf{v}'}{dt'}$ and $\displaystyle\textbf{a}_{u}=\frac{d\textbf{u}}{dt}$. The equations of motion, deduced by Newton, are valid in the fixed system $O'$, by construction of his theory. So, a particle with mass $m$ subjected to a force $\textbf{F}$ $=$ $\textbf{F}(\textbf{r}, \textbf{v}, t)$, obeys
\begin{eqnarray}
 m\frac{d^{2}\textbf{r}}{dt^{2}} = \textbf{F}(\textbf{r}, \textbf{v}, t),
 \label{N}
\end{eqnarray}
in the $O$ referential. Making use of the eq. (\ref{adacel}) you can write the equation of motion in $O'$ as
\begin{eqnarray}
 m\frac{d^{2}\textbf{r}'}{dt'^{2}} = \textbf{F}'(\textbf{r}', \textbf{v}, t') - m\textbf{a}_{u}.
 \label{NewtonO'}
\end{eqnarray}
The term m$\textbf{a}_{u}$ cannot be interpreted as a force in the physical sense, but only the product of mass by the acceleration of $O'$ relative to $O$. In fact, this product is called \textit{pseudo-force} or \textit{fictitious force}. From point of view of Classical Mechanics, $-m\textbf{a}_{u}$ not a force, because it's not due to the physical phenomenon of the nature \cite{Symon}.

If $O'$ move with constant speed relative to $O$, so $\textbf{a}_{u}$ $=$ $\textbf{0}$ and with that, the equation (\ref{NewtonO'}) is
\begin{eqnarray}
    m\frac{d^2\mathbf{r}'}{dt'^2} = \mathbf{F}'(\mathbf{r}', \mathbf{v}',t'),
\label{N'}
\end{eqnarray}
Thus, showing that the Newton's equations of motion are valid in any reference system that moves relative to another frame with constant velocity. Such reference systems are called \textit{inertial frames} and this covariance between the equation of motion (\ref{N}) and (\ref{N'}) translate the then known \textit{Newtonian Principle of Relativity}, that can be enunciated as follows:\\

\textit{"The Newton's laws are the same in all inertial frame and, with effect, \textbf{no mechanical experiment} is able to reveal to the observer information about the motion of an inertial frame in relation to another inertial frames"}\\

This principle affirms that there is no a privileged inertial frame, that is, they are all equivalent, making it impossible a absolute frame of reference. We can raise the following questions about how the position measures between two inertial frames $O$ and $O'$ are related. We can consider here that the inertial observers are using instruments calibrated and synchronized with each other, that so lengths and time intervals are absolute. Solving $\displaystyle \frac{d^{2}\textbf{r}}{dt^{2}}$ $=$ $\textbf{0}$ and admit that the two frames, $O$ and $O'$, coincide in $t=0$, we have
\begin{eqnarray}
    \frac{d^{2}\textbf{h}}{dt^{2}} = \frac{d\textbf{u}}{dt} = \textbf{0} \Rightarrow \textbf{u} = \textbf{constant} \Rightarrow \textbf{h}(t) = \textbf{u}t
\label{h(t)}
\end{eqnarray}
replacing (\ref{h(t)}) in (\ref{r(t)}), we have to 
\begin{eqnarray}
   \textbf{r}' = \textbf{r} - \textbf{u}t.
    \label{r'}
\end{eqnarray}
Joining the equation (\ref{r'}) with the fundamental hypothesis of Classical Mechanics
\begin{eqnarray}
    \textbf{r}' = \textbf{r} - \textbf{u}t,\quad t'= t,
\label{TG}
\end{eqnarray}
These equations are the \textit{Galileo's transformations}. There is a special case of the Galileo's transformations (TG), when the movement of $O'$ is along to $x$-axes of $O$, Fig. \ref{referenciais} (b). In this case, the equation (\ref{TG}) is in the form
\begin{eqnarray}
    x'=x - ut, \quad y'=y, \quad z'=z, \quad t'=t.
\label{TGpadrao}
\end{eqnarray}
This set of equations are called \textit{Galileo's transformation into standard form}.

\section{Review of Fundamental Concepts of Equilibrium Thermodynamics}
\label{Termo}

The thermodynamics is a branch of Physics that studies the relation between the energy transformations and work involving thermal phenomena. Thus, it's necessary to define precisely some fundamental concepts \cite{Callen}. A \textit{system} is a limited region of space or a portion of finite matter on which an analysis is performed. The \textit{neighborhood} of a thermodynamic system is everything that is outside the system and is separated from the same system by its \ textit {frontier}, which can be real or imaginary, fixed or mobile. The macroscopic quantities related to the state of a system are called \textit{macrostates} and we will be interested in systems whose transformations from one macrostate to another are done stationary, that is, \ textit {in equilibrium}. This simplification becomes necessary so that thermodynamic systems can be fully characterized by three extensive parameters, namely: \textit{internal energy} $U$, \textit{volume} $V$ and \textit{mole numbers} $N$. So, any macrostate of a thermodynamic systems in equilibrium are characterized by the set $(U,V,N)$.

The simplest criterion for determining the final equilibrium state of a thermodynamic system is to define a extensive function of the parameters $(U,V,N)$ called \textit{entropy}, defined as a continuous and differentiable function $S=S(U,V,N)$ with respect to all extensive parameters. The entropy of a simple system is a \textit{homogeneous first-order function} of the extensive parameters, i.e. all $\lambda$ real, we have
\begin{eqnarray}
S(\lambda U,\lambda V,\lambda N)=\lambda S(U,V,N).
\label{homogeneidade}
\end{eqnarray}
Another property of the entropy of simple thermodynamic systems is the \textit{additivity}. If $S_i(U,V,N)$ is the entropy of a $i$th subsystem, so the entropy of complete system will be  
\begin{eqnarray}
S(U,V,N)=\sum_iS_i(U,V,N).
\end{eqnarray}
The last fundamental property of entropy says that it is 
\textit{increases monotonically} with the internal energy $U$, that is
\begin{eqnarray}
\left(\frac{\partial S}{\partial U}\right)>0.
\end{eqnarray}
The continuity, differentiability and monotonicity of entropy imply that it can be inverted (Implicit Function Theorem, \cite{Spivak}) with respect to internal energy, that is $S=S(U,V,N)\Rightarrow U=U(S,V,N)$, where $U$ is also univocal, continuous and differentiable function. If $U$ is taken as the fundamental relation, Thermodynamics can b built on the \textit{energy representation}, taking the total differential of the internal energy
\begin{eqnarray}
dU=\left(\frac{\partial U}{\partial S}\right)dS+\left(\frac{\partial U}{\partial V}\right)dV+\left(\frac{\partial U}{\partial N}\right)dN,
\end{eqnarray}
and with this equation we can defined the thermodynamic \textit{state equations} \cite{Callen}:
\begin{eqnarray}
T=\frac{\partial U}{\partial S}, \quad p=-\frac{\partial U}{\partial V}, \quad \mu=\frac{\partial U}{\partial N},
\label{eqsEstado}
\end{eqnarray}
where $T$ is the absolute temperature, $p$ is pressure and $\mu$ is the chemical potential. Replacing the state equations in the differential $dU$ we get the \textit{fundamental relation of Thermodynamic in the energy representation}
\begin{eqnarray}
dU=TdS-pdV+\mu dN.
\label{RFT}
\end{eqnarray}
In the entropy representation, the previous equation can be written as
\begin{eqnarray}
dS=\frac{dU}{T}+\frac{p}{T}dV-\frac{\mu}{T}dN.
\end{eqnarray}
From the relation (\ref{RFT}), considering an infinitesimal transformation from one microstate to another, with fixed mole numbers $(dN=0)$, one arrives at \textit{First Law of Thermodynamics}:
\begin{eqnarray}
dQ=dU-dW,
\end{eqnarray}
where $dW=-pdV$ is the work realized on the system and $\displaystyle dS=\frac{dQ}{T}$  is the amount of heat transferred or absorbed in the \textit{reversible} process. With that you can advertise the \textit{Second Law of Thermodynamics}, as \cite{Greiner}

\textit{"The variation of entropy of a thermodynamically isolated system is null for reversible process and is always greater than zero for irreversible process. Mathematically,}
\begin{eqnarray}
\Delta S\geqslant0.
\end{eqnarray}
The second law of thermodynamics affirm is that, in an irreversible process, the system searches for a new equilibrium state, so that the entropy of the system grows until it reaches a maximum value in equilibrium. For the reversible processes, the entropy does not grow during the process. 

\section{Transformation Between Different Referential frames in Thermodynamics}
\label{Transf}

Now we'll look at what happens to the laws and the thermodynamic quantities when viewed from non-inertial observer. First, we must analyze the mechanical work, when this is calculated by non-inertial observer. Here is convenient to work with a large particle numbers, as it is simpler to relate between mechanical quantities with thermodynamic quantities.

\subsection{Microscopic analysis}

Consider the analysis of mechanical work on transformation between non-inertial frames. Assuming that the system under study is composed of $N$ particles subjected to external forces, which act on the system by ith particle due to $N-1$ remaining particles belonging to the system, so that the Newton's second law (\ref{NewtonO'}), in the $O'$ referential becomes 
\begin{eqnarray}
m_i\frac{d\mathbf{v}'_i}{dt'}=\mathbf{F}_i'+\sum_{j=1}^N\mathbf{F}'_{ij}(\mathbf{r}'_{ij})-m_i\mathbf{a}_h.
\end{eqnarray}
Calculating the scalar product of this equation with $\mathbf{v}'_i$ and adding over all particles, we have
\begin{eqnarray}
dK'=dW'+dW_{int}+dW_h,
\label{dK'}
\end{eqnarray}
where 
\begin{eqnarray}
K'&=&\sum_{i=1}^{N} \frac{m_{i} v_{i}^{'2}}{2},\\
dW'&=& \sum_{i=1}^{N} \mathbf{F}^{'}_{i} \cdot d\mathbf{r}^{'}_{i},\\
dW_{int}&= &\sum_{i=1}^{N}\sum_{i<j=1}^{N}\mathbf{F}_{ij}\cdot d\mathbf{r}_{ij}, \label{Wint}\\
dW_{h}&=& -\left(\sum_{i=1}^{N}m_{i}\mathbf{v}^{'}_{i}\right)\cdot d\mathbf{u}, \label{dWhMi}
\end{eqnarray}
are the total kinetic energy, the amount of infinitesimal work of the external forces, internal to the referential $O'$ and the infinitesimal work of pseudo force, respectively, being $\mathbf{r}'_{ij}:=\mathbf{r}'_i-\mathbf{r}'_j=\mathbf{r}_i-\mathbf{r}_j=\mathbf{r}_{ij}$. The meaning of $dW_h$ is the infinitesimal fictional work done on the system.

Considering the conservative internal forces, we can write them as the negative of a gradient of an internal potential energy \cite{Symon}, calculated with respect to the relative coordinates $\mathbf{r}_{ij}$, causing (\ref{Wint}) stay written as 
\begin{eqnarray}
dW_{int}=-\sum_{i=1}^N\sum_{i<j=1}^N\nabla_{ij}V_{(int)}\cdot d\mathbf{r}_{ij}= - dV_{int},
\end{eqnarray}
In which can be define the \textit{internal energy} or \textit{own energy} of the system at the referential $O'$ as 
\begin{eqnarray}
U'= K'+V_{(int)},
\end{eqnarray}
and so the equation (\ref{dK'}), becomes
\begin{eqnarray}
dU'=dW'+dW_h.
\label{dU'}
\end{eqnarray}
In this context, the only way to introduce the heat, measured in relation to the referential $O'$ is to add on term that represents it in the right member of the equation (\ref{dU'}). Soon we have
\begin{eqnarray}
dQ'=dU'-dW'-dW_h.
\label{dU'}
\end{eqnarray}
It is the observed, that the first law is only invariant in cases where $dw_h=0$. The cases of physical interest, where this fact occurs are two: (i) Galileo's inertial frames $\displaystyle \left(\mathbf{a}_h=0\right)$ and (ii) when the referential $O'$ is the center of mass $\displaystyle \left(\sum_{i=1}^Nm_i\mathbf{v}'_i=\mathbf{0}\right)$.

The analysis of entropy transformation in the microscopic context if done through of the entropy defined by Boltzmann \cite{Reif}
\begin{eqnarray}
  S(U,V,N)=k_B\ln\Omega(U,V,N),
  \label{S}
\end{eqnarray}
where $\Omega(U,V,N)$ is the number of accessible microstates to the thermodynamic system and $k_B$ is the Boltzmann's constant. According to equation (\ref{S}), to determine the entropy transformation between the reference frames $O$ and $O'$, will depend on the transformation of the quantities $U,$ $V$ and $N$. The volume and mole numbers are identities transformations ($V'=V$ e $N'=N$), because in the Newtonian relativity don't exist contraction of volume or change of particle numbers. Considering that the mass center of the particles system is found in rest relative to the $O'$ frame. we can be shown that the internal energy is transformed as
\begin{eqnarray}
  U'(S,V,N,\mathbf{u})=U(S,V,N)-\frac{Mu^2}{2},
 \label{U'}
\end{eqnarray}
where $M$ is the total mass of the particles system. To build this form, we can write
\begin{eqnarray}
  \Omega (U,V,N)&=&\Omega (U'+Mu^2/2,V',N')\nonumber\\
                &=&\Omega' (U',V',N'),
\end{eqnarray}
where $\Omega' (U',V',N')$ is the count of the microstates in the $O'$ frame. So, immediately we can see that the \textit{entropy given by the equation (\ref{S}) is invariant on transformation between accelerated reference frames.}
\subsection{Macroscopic analysis}

In the section \ref{Termo}, it was shown that an simple thermodynamic system, in equilibrium, is fully characterized by energy $U$, volume $V$ and particles number $N$. We now want to conduct an analysis of the point of view of accelerated moving references in a \textit{macroscopic} way. Now, consider that an arbitrary state of system in relative motion are fully determined, besides $U'$, $V'$ e $N'$, by relative velocity between the reference frames, $\mathbf{u}$. In this case, the fundamental relation, in the energy representation (\ref{RFT}), at the accelerated frame, can be obtained from
\begin{eqnarray*}
dU'=\left(\frac{\partial U'}{\partial S'}\right)dS'+\left(\frac{\partial U'}{\partial V'}\right)dV'+\left(\frac{\partial U'}{\partial N'}\right)dN'+ \nabla_{\mathbf{u}}U'\cdot d\mathbf{u}.
\end{eqnarray*}

Using the definitions (\ref{eqsEstado}) and defining a new set of intensive parameters, associated to motion of $O'$ in relation to $O$, we can defined
\begin{eqnarray}
\mathbf{P}'= -\nabla_{\mathbf{u}}U',
\label{P'}
\end{eqnarray}
where $\nabla_{\mathbf{u}}U'$ is a gradient of $U'$ relative to velocity coordinates $\mathbf{u}$, defined by 
\begin{eqnarray}
\nabla_{\mathbf{u}}U'=\frac{\partial U'}{\partial u_x}+\frac{\partial U'}{\partial u_y}+\frac{\partial U'}{\partial u_z}.
\end{eqnarray}
Finally, the differential equation of $U'$ in the $O'$ frame is
\begin{eqnarray}
  dU'=T'dS'-p'dV'+\mu' dN'-\mathbf{P'}\cdot d\mathbf{u}.
  \label{RFT'}
\end{eqnarray}
In this case, the heat provided, $dQ'$, is equal to the increase of total energy minus the produced work by pressure and that is used in the increase of momentum, so that the \textit{first law of Thermodynamics in the frame of reference} $O'$, become (for $dN'=0$)
\begin{eqnarray}
dQ'=dU'-dW'-dW_h,
\label{dQ'}
\end{eqnarray}
where
\begin{eqnarray}
dW_h=-\mathbf{P}'\cdot d\mathbf{u}.
\label{dWhMa}
\end{eqnarray}
To compare the equations (\ref{dWhMi}) and (\ref{dWhMa}), we can conclude that the vector parameter $\mathbf{P}'$ is the \textit{ total linear momentum of the particles that constitute the matter in relation to the reference frame O'}, because each velocity $\mathbf{v}'_i$ is calculated based on $O'$, accordingly to equation (\ref{advel}).

We now consider a thermodynamic system in some internal state of interest, however, at rest and with defined entropy $S$. We will then accelerate it in a reversible and adiabatic way without changing the internal state of system, so that the entropy some moment of specific time are $S'$. In this way, as we realize such procedure of adiabatic form
\begin{eqnarray}
\Delta S=S'-S=\int \frac{dQ}{T}=0 \Rightarrow S=S',
\end{eqnarray}
so that, \textit{the entropy is an invariant quantity under transformations between accelerated reference frames}, thus agreeing with the microscopic analysis. So, taking into account this invariance of the entropy, as well the invariance of volume ($V'=V$) and particle number ($N'=N$), assumed to be true in Classical Mechanics, the relation (\ref{RFT'}) is
\begin{eqnarray}
  dU'=T'dS-p'dV+\mu' dN -\mathbf{P'}\cdot d\mathbf{u}.
  \label{RFT'2}
\end{eqnarray}
Here we cannot say that $U'=U$ because, mathematticaly, $U=f(S,V,N)$ and $U'=g(S,V,N,\mathbf{u})$, showing that $U'\neq U$. Equality can be only assumed in the case where $\mathbf{u}=\mathbf{0}$, so that, when we are considering fixed reference frames.  

\subsection{Transformations between Thermodynamic Quantities}
\label{transformação}

At this stage, we must determine the relations between temperatures, pressures and chemical potentials measured by observers $O$ and $O'$. For this, we will show the invariance of entropy, volume and mole number will be of fundamental importance. 

First consider the analysis of the absolute temperature measured by the different reference frames $O$ and $O'$. Let $X_i$ be the extensive thermodynamic parameters $S$, $V$ and $N$ ($X_1 = S$, $X_2 = V$ and $X_3 = N$) and $Y_i$ the intensive thermodynamic parameters $T$, $p$ and $\mu$ ($Y_1 = T$, $Y_2 = p$ and $Y_3 = \mu$). In this way the state equations (\ref{eqsEstado}) can be written compactly as $\displaystyle Y_i = \frac{\partial U}{\partial X_i}$. These state equations in the $ O '$ referential, combined with the equation (\ref{U'}), provide
\begin{eqnarray}
Y_i'=\frac{\partial U'}{\partial X_i'}=Y_i-M\mathbf{u}\cdot\frac{\partial \mathbf{u}}{\partial S}.
\label{Yi}  
\end{eqnarray}
We must find a functional relation between velocity $\mathbf{u}$ and entropy $S$. 
However, as $\mathbf{u}$ is a unique function of time $t$, we can see that $\partial \mathbf{u}/\partial S=0$. Then, it concludes that \textit{the equations of state are invariant under any change of reference frames in translation}. Since the pressure $p$ and volume $V$ are invariant under coordinates transformation between accelerated frames, so
\begin{eqnarray}
dW'=-p'dV'=-pdV=dW,
\end{eqnarray}
showing that \textit{the work $dw$ of expansion or compression carried out on the thermodynamic system is invariant under any transformation of translation between reference frames}. On the other hand, the internal energy in the $O'$ reference obeys the equation (\ref{U'}). Then, 
\begin{eqnarray}
dU'=dU-M\mathbf{u}\cdot d\mathbf{u}.
\label{dU'}
\end{eqnarray}
We see clearly that the \textit{variation of internal energy in the frame $O'$ is not invariant under transformation of translation between reference frames, except in cases where the frames are inertial}, because the latter $\mathbf{u}$ is constant, causing $d\mathbf{u}=0$. According with the equations (\ref{P'}) and (\ref{dWhMa}), we have $\mathbf{P}'=M\mathbf{u}$, and we can write (\ref{dU'}) as
\begin{eqnarray}
dU'=dU+dW_h.
\end{eqnarray}
When replacing the last equation in the First Law (\ref{dQ'}), we see that the transformation of $dQ'$ will be
\begin{eqnarray}
dQ'=dU-dW=dQ.
\label{dQ'=dQ}
\end{eqnarray}
showing that the \textit{amount of heat $dQ$ also is invariant under any transformation of translation between reference frames}. Another way to analyze this result would consider reversible systems, which $dS=dQ/T$. As entropy and temperature are invariant, the amount of heat $dQ$ will also be.

\section{Applications}
\label{Aplic}

As immediate applications of Thermodynamic Theory between non-inertial reference frames exposed so far, let's consider two classic problems addressed in fixed references: (i) the problem of ideal gas such as the equation of state and Halley's law (pressure varying with height) and (ii) Carnot's cycle and thermal machines, to determine the invariance or non-invariance of the efficiency of this type of machine.

\subsection{The ideal gas}
\label{Gás}

Consider the analysis of some properties of the equation state of a ideal gas when a observer $O'$ in translation with relation to a second observer $O$ conducts an analysis of a sample of ideal gas. The equation of state of an ideal gas is given by famous relation $pV=nRT$, where $n=Nk_B$ \cite{Callen}. However, according with the section \ref{transformação}, explored previously, we can see that this equation is invariant under classic spatial translation, because both temperature, pressure and volume are invariant under transformation of translation between reference frames. However, we can analyze how a moving observer measures the pressure as function of height in it proper frame of reference, known as Halley's law \cite{Moyses, Alonso}. First, we must know how it looks the law of static equilibrium for fluids in moving references. The Euler's equation, which provides the fluid dynamics in motion, at the fixed frame $O$ is given by \cite{Landau2} 
\begin{eqnarray}
\rho\frac{d\mathbf{v}}{dt}+\nabla p=\mathbf{f},
\end{eqnarray}
where $\mathbf{f}$ is volumetric force that acts on the fluid. Replacing (\ref{advel}) in the previous equation, we see that in the frame $O'$, the Euler equation is
\begin{eqnarray}
\rho\frac{d\mathbf{v}'}{dt}+\sum_{i=1}^3\sum_{j=1}^3\frac{\partial x'_j}{\partial x_i}(\nabla p)_j\mathbf{\hat{e}}_i=\mathbf{f}-\rho\mathbf{a}_h.
\label{Euler'}
\end{eqnarray}
For the cases where $\displaystyle \frac{\partial x'_j}{\partial x_i}=\delta_{ij}$, being $\delta_{ij}$ the Kronecker's delta \cite{Arfken}, we see that $\nabla p=\nabla'p$. So, the equation (\ref{Euler'}) becomes
\begin{eqnarray}
\rho\frac{d\mathbf{v}'}{dt}+\nabla' p=\mathbf{f}-\rho\mathbf{a}_h.
\end{eqnarray}
If the fluid is in equilibrium in relation to observer $O'$, must have $\displaystyle \frac{d\mathbf{v}'}{dt}=\mathbf{0}$, resulting in
\begin{eqnarray}
\mathbf{f}=\nabla' p+\rho\mathbf{a}_h.
\label{f}
\end{eqnarray}
If the volumetric force is the is weight force, so $\mathbf{f}=\rho\mathbf{g}$. Then, we can write (\ref{f}) as
\begin{eqnarray}
-\rho g\mathbf{\hat{z}}=\nabla' p+\rho\ddot{\mathbf{h}},
\end{eqnarray}
so that
\begin{eqnarray}
\frac{\partial p}{\partial z'}=-\rho(g+\ddot{h}_z), \quad \frac{\partial p}{\partial x_i'}=-\rho\ddot{h}_i, \quad i=1,2.
\label{pz}
\end{eqnarray}
Knowing that $\displaystyle \frac{\partial p}{\partial z'}=-\rho g$ \cite{Moyses, Alonso}, we can write the first equation (\ref{pz})
\begin{eqnarray}
\frac{\partial p}{\partial z'}=\frac{\partial p}{\partial z}-\rho\ddot{h}_z.
\label{dp}
\end{eqnarray}
Note that, like $\displaystyle \frac{\partial p}{\partial z}<0$, the signal of $\displaystyle \frac{\partial p}{\partial z}$ will depends of signal of $\ddot{h}_z$.
Two cases of particular interest occurs when the $O'$ frame is inertial and when it is in free fall. Considering initially the case in that the $O'$ frame is inertial, so $\ddot{\mathbf{h}}=\mathbf{0}$. Thus, it should be noted that the equilibrium equation (\ref{f}) and (\ref{dp}) become invariant,
\begin{eqnarray}
\mathbf{f}=\nabla' p=\nabla' p, \quad \frac{\partial p}{\partial z'}=\frac{\partial p}{\partial z}=-\rho g.
\label{inertial}
\end{eqnarray}
Another case is when the frame of reference is free fall, choosing $\ddot{\mathbf{h}}=\mathbf{g}=-g\hat{\mathbf{z}}$. Replacing this equation in (\ref{pz}), we see that
\begin{eqnarray}
\frac{\partial p}{\partial z'}=0.
\end{eqnarray}
This result informs that the observer, in the proper frame of reference $O'$, do not detect any changing of the pressure relative to weight $z$. Physically, this result is expected, since the measure of weight that the observer $O'$ makes is always null, therefore, the pressure cannot change. It is inevitable not to compare this result with Einstein's principle of equivalence \cite{Weinberg}.

\subsection{Carnot's Cycle}

The problem of Carnot's cycle is based on the concept of \textit{thermal engine}. This device is a machine it produces work from heat, operating cyclically obeying the second law of Thermodynamics as stated by Clausius \cite{Halliday}, that states that it is impossible to realize a process such that only effect is removing heat of a hot thermal reservoir and producing an amount of work. Then, we can formulate the problem of Carnot's cycle as follow: given a hot and cold sources, what the maximum efficiency, through mechanical work, which can be obtained of a thermal engine operating in cycles between these two sources? The efficiency of Carnot's thermal machine is defined as \cite{Callen}
\begin{eqnarray}
\eta=-\frac{W}{Q_2}=1-\frac{Q_1}{Q_2}=1-\frac{T_1}{T_2},
\label{eta}
\end{eqnarray}
where $Q_2=-W+Q_1$ is the heat provided by hot source, $Q_1$ is the heat absorbed by cold source and $W$ is the mechanical work realized in the process during one cycle. Here we desire to know if the efficiency of a Carnot's machine is affected by change of observer, i.e., if it changes with the observer moving it's moving in relation to Carnot's machine. For this, we will simply use the invariance of amount of heat, i.e, $Q_1=Q'_1$ e $Q_2=Q'_2$ in (\ref{eta})
\begin{eqnarray}
\eta=1-\frac{Q'_1}{Q'_2}=\eta',
\label{eta=eta'}
\end{eqnarray}
showing that the \textit{performance of Carnot's machine is also an invariant under any transformation of translation between frams of reference}. This is physically expected, because an thermal machine cannot increase its efficiency due the motion of a reference, if not the car engine, for example, would have two different efficiencies: one in rest relative to the ground and another when it was in motion, , which is not physically observed.

\section{Conclusion}

From understanding of relative motion in Newtonian Mechanics it was possible to establish the relation between measures of the thermodynamic parameter in different reference frames. Using the concept of translation motion between frames $O$ and $O'$, it was possible to apply in the thermodynamic context for extracting important results, that are not addressed in the textbooks on the subject, among them the problem of ideal gas and Carnot's cycle.

First we show that the \textit{First Law of Thermodynamics} is invariant under any transformation of spatial translation between reference frames. This can be seen by the heat transformation equation (\ref{dQ'=dQ}). However, we show that the entropy is a invariant under any change of frame in translation relative to another fixed, it showing that the condition of accelerated or inertial frame of reference does not influence a chance of entropy. Consequently, the \textit{Second Law of Thermodynamic will be invariant}. Still in the theoretical development, we prove that the state equations also are covariant under any transformation of translation between inertial reference frames, because $\mathbf{u}$, in the equation (\ref{Yi}), is an explicit function of time only and not of entropy. This fact leads to invariance of classic state equation of an ideal gas $pV=nRT$.

After theoretical development, we were done two applications, that was about the pressure variation with height and the problem of Carnot's machine. In the first, we address the problem of a general point of view and after we highlight two cases: (i) inertial frames and (ii) frames in free fall. For the first case, we show that the equation that provides the law of pressure variation with height remains invariant, according with the equations (\ref{inertial}), since $\mathbf{\ddot{h}}=\mathbf{0}$. For the second case, we insert $\ddot{\mathbf{h}}=\mathbf{g}$ in the equation (\ref{dp}) and verify that the observer in the frame of reference in free fall does not detect pressure changes, because in it's own frame it always is in the same height. In the analysis of the Carnot's machine, was given attention to it performance. We show that efficiency of this machine does not depend only on the amounts of heat or temperature of hot and cold sources. As these quantities are invariant under spatial translation between reference frames, the performance of the machine is also invariant, according to the equation (\ref{eta=eta'}).

\bibliography{Template} 

\begin{thebibliography}{10}
\newcommand{\enquote}[1]{``#1''}

\bibitem{Noel}
Noel Doughty, \emph{Lagrangian interaction: an introduction to relativistic
  symmetry in electrodynamics and gravitation} (CRC Press, 2018).

\bibitem{Werner}
Werner Heisenberg, \emph{Physics and beyond} (Allen \& Unwin London, 1971).

\bibitem{Callen}
Herbert~B Callen, \enquote{Thermodynamics and an introduction to
  thermostatistics,}  (1998).

\bibitem{Tykodi}
Ralph~J Tykodi, \enquote{Thermodynamics and classical relativity,} American
  Journal of Physics \textbf{35}~(3), 250--253 (1967).

\bibitem{Van}
NG~Van~Kampen, \enquote{Relativistic thermodynamics of moving systems,}
  Physical Review \textbf{173}~(1), 295 (1968).

\bibitem{Muller}
I~M{\"u}ller and I~Liu, \enquote{Extended thermodynamics of classical and
  degenerate gases,} Arch. Rat. Mech. Anal \textbf{83}~(4), 285--332 (1983).

\bibitem{Ruggeri}
T~Ruggeri, \enquote{Galilean invariance and entropy principle for systems of
  balance laws,} Continuum Mechanics and Thermodynamics \textbf{1}~(1), 3--20
  (1989).

\bibitem{Symon}
KR~Symon, \enquote{Mechanics addison-wesley,} Reading, MA \textbf{1}, 971
  (1971).

\bibitem{Landau}
Lev~Davidovich Landau and EM~Lifschic, \emph{Course of theoretical physics.
  vol. 1: Mechanics} (Oxford, 1978).

\bibitem{Costas}
Costas~J Papachristou, \emph{Introduction to Mechanics of Particles and
  Systems} (Springer Nature, 2020).

\bibitem{Tung}
Wu-Ki Tung, \emph{Group theory in physics}, volume~1 (World Scientific, 1985).

\bibitem{Jerrold}
Jerrold~E Marsden and Tudor~S Ratiu, \emph{Introduction to mechanics and
  symmetry: a basic exposition of classical mechanical systems}, volume~17
  (Springer Science \& Business Media, 2013).

\bibitem{Spivak}
Michael Spivak, \emph{Calculus on manifolds: a modern approach to classical
  theorems of advanced calculus} (CRC press, 2018).

\bibitem{Greiner}
Walter Greiner, Ludwig Neise, and Horst St{\"o}cker, \emph{Thermodynamics and
  statistical mechanics} (Springer Science \& Business Media, 2012).

\bibitem{Moyses}
Herch~Moys{\'e}s Nussenzveig, \emph{Curso de F{\'\i}sica B{\'a}sica: fluidos,
  oscila{\c{c}}{\~o}es e ondas, calor}, volume~2 (Editora Blucher, 2018).

\bibitem{Alonso}
Marcelo Alonso and Edward~J Finn, \emph{Fundamental university physics},
  volume~2 (Addison-Wesley Reading, MA, 1967).

\bibitem{Landau2}
Lev~Davidovich Landau and Evgenii~Mikhailovich Lifshits, \emph{Fluid mechanics,
  by LD Landau and EM Lifshitz}, volume~11 (Pergamon Press Oxford, UK, 1959).

\bibitem{Arfken}
Hans~J Weber and George~B Arfken, \emph{Essential mathematical methods for
  physicists, ISE} (Elsevier, 2003).

\bibitem{Weinberg}
Steven Weinberg, \enquote{Gravitation and cosmology: principles and
  applications of the general theory of relativity,}   (1972).

\bibitem{Halliday}
David Halliday, Robert Resnick, Kenneth~S Krane, and Paul Stanley,
  \emph{Physics, Volume 1}, volume~1 (2001).

\end{thebibliography}







\end{document}